\documentclass[twocolumn,prl,aps,amsmath,amssymb,superscriptaddress,showpacs]{revtex4}
\usepackage{dcolumn}
\usepackage{hyperref}
\usepackage{multirow}
\hypersetup{
    colorlinks=true,
    linkcolor=blue,
    filecolor=magenta,      
    urlcolor=black,
    citecolor=blue,
}
\makeatletter
\g@addto@macro\normalsize{%
  \setlength\abovedisplayskip{12pt}
  \setlength\belowdisplayskip{12pt}
  \setlength\abovedisplayshortskip{12pt}
  \setlength\belowdisplayshortskip{12pt}
}
\makeatother
\usepackage{physics}
\usepackage{epsfig}
\usepackage{graphics}
\usepackage{float}
\usepackage{amsmath}
\usepackage{xcolor,colortbl}
\usepackage[utf8]{inputenc}
\usepackage[T1]{fontenc} 
\usepackage[colorinlistoftodos]{todonotes}
\usepackage{lmodern} 
\usepackage{soul}
\usepackage[version=4]{mhchem}
\usepackage{color}
\usepackage{soul}

\definecolor{Gray}{rgb}{0.72,0.72,0.98}
\definecolor{LightCyan1}{rgb}{0.83,0.83,0.98}
\definecolor{LightCyan2}{rgb}{0.91,0.92,1}

\begin{document}

\title{Optical properties of \ce{TiS3} as a novel thin film for single-junction and tandem solar cells}

\author{Cesar E. P. Villegas}
\affiliation{ Departamento de Ciencias, Universidad Privada del Norte, Lima 15434, Peru} 
\author{Enesio Marinho Jr}
\affiliation{
 Departamento de Física e Química, Faculdade de Engenharia, Universidade Estadual Paulista (UNESP), Av. Brasil 56, Ilha Solteira, São Paulo 15385-007, Brazil}
\author{A. C. Dias}
\affiliation{Instituto de Física e Centro Internacional de Física, Universidade de Brasília, Brasília-DF 70919-970, Brazil}
\author{Pedro Venezuela}
\affiliation{Instituto de Física, Universidade Federal Fluminense (UFF),\\ Av.\ Gal.\ Milton Tavares de Souza, s/n, 24210-346 Niterói, Rio de Janeiro, Brazil.}

\author{Alexandre R. Rocha}
\affiliation{
 Instituto de Física Teórica, Universidade Estadual Paulista (UNESP),\\
R.\ Dr.\ Bento Teobaldo Ferraz, 271, São Paulo, 01140-070 São Paulo, Brazil.}


\begin{abstract}
Sub-micrometer thin films are promising platforms for emerging flexible photovoltaic devices. Although the current market already produces efficient solar cells, the average wafer thickness of these devices remains far from the sub-micrometer scale, making them susceptible to cracking under bending stress and thus precluding their use in flexible device applications.
Due to its earth abundance, non-toxicity, and low elastic modulus, titanium trisulfide (\ce{TiS3}) has emerged as a promising alternative for flexible device applications. Here, using excited-state density functional calculations combined with the transfer matrix approach, we perform an optical analysis and assess the efficiency of a prototype photovoltaic device based on sub-micrometer \ce{TiS3} thin films.
Using optical constants obtained from our first-principles calculations, we evaluate the photovoltaic response of a single-junction device in the radiative limit, finding that a 140-nm-thick active layer achieves a maximum power conversion efficiency of approximately 22\%. Additionally, we investigate tandem solar cells that incorporate \ce{TiS3} into perovskite thin films, and find that the lower and upper power conversion efficiencies range from approximately 18\% to 33\%.
Overall, our results suggest great potential for using \ce{TiS3} thin films as an active layer in the design of highly efficient flexible solar cells.
\end{abstract}

\maketitle

\pagebreak 

\section{Introduction}
Single-junction solar cells, devices based on a single absorber, are currently the primary technology in the photovoltaic industry. These devices, which commonly include crystalline silicon (c-Si) and thin-film technologies based on GaAs, CdTe, Cu(In,Ga)Se$_{2}$ (CIGS), and perovskites have undergone a significant increase in efficiency over time \cite{national2023best}, reaching power conversion efficiencies (PCEs) above 22$\%$ \cite{green2023solar} and, in some cases,  approaching the Shockley–Queisser limit \cite{shockley1961detailed,ruhle2016tabulated}. This is the case for GaAs single-junction solar cells,  which have achieved a record efficiency of $29.1\%$ with only a few microns of absorber material \cite{green2023solar,kayes201127}, or c-Si, which has achieved a PCE of 26.8$\%$ for a 165-$\mu$m-thick active layer \cite{green2023solar,yoshikawa2017silicon}. Although silicon-based photovoltaic cells dominate the solar cell market, with a market share of roughly 95$\%$ \cite{razzaq2022silicon}, they typically possess an average wafer thickness of 170 $\mu$m \cite{ballif2022status}, making them not only relatively heavy but also rigid and susceptible to cracking under bending stress \cite{liu2023flexible,tachibana2023development}, which ultimately precludes their use for flexible device applications.

Reducing the absorber's thickness in a single-junction solar cell provides device flexibility while reducing manufacturing costs \cite{liu2023flexible,li2024flexible}. Moreover, it is also a successful route to mitigate non-radiative bulk recombination processes \cite{rai2020effect,luo2020minimizing}, enabling the use of absorber materials with reduced diffusion lengths. Thus, if successful, thin-film-based solar cell technologies are expected to be integrated into emergent portable energy storage systems, as well as flexible and wearable electronic devices \cite{hashemi2020recent}.
Currently, thin-film solar cells based on CIGS, CdTe, and perovskites present the most notable features regarding flexibility, electronic compatibility and efficiency. Particularly, perovskite cells can operate with efficiencies closer to 25$\%$ \cite{green2023solar,ramanujam2020flexible,jung2019flexible}. Nonetheless, the scarcity of chemical elements in the earth's crust, toxicity, and stability issues of these emergent thin-films still restricts their widespread commercialization. Thus, investigating abundant, non-toxic absorber materials with optimal electronic and optical properties is crucial for the design and development of next-generation photovoltaic devices \cite{alharbi2011abundant}.

An alternative way to significantly improve photovoltaic device efficiency is the design of tandem solar cells, which typically consist of a stacked arrangement of two or more absorber materials  \cite{marti1996limiting}. Regarding tandem devices based on two absorbers, the top layer usually has a larger bandgap (around 1.5--1.7 eV), while the bottom layer presents a smaller one (between 0.9--1.2 eV) \cite{eperon2017metal}. Hence, in principle, the top cell allows the efficient absorption of photons with higher energies, which minimizes thermalization losses, enabling the transmission of photons with energies in the near-infrared spectrum to be absorbed by the bottom cell. By employing this solar cell setup, perovskite/CIGS tandem devices with PCEs of 24.2$\%$ have been achieved in single crystals \cite{jost2022perovskite}. More recently, LONGi Green Energy Technology Co. announced a new world record efficiency of 34.6$\%$ for two-absorber tandem solar cells \cite{longi2024}.

Titanium (Ti) and sulfur (S) are relatively abundant in the earth's crust \cite{hayneshandbook2018} and non-toxic elements that can form stable binary compounds such as titanium trisulfide (\ce{TiS3}) \cite{tripathi2021review}, which is a semiconductor with an optical band gap of $\sim$1 eV \cite{ferrer2013optical,molina2015electronic} that closely approximates the value predicted by the Shockley-Queisser limit for maximum photon-to-electrical current conversion. Therefore it can, potentially, be used as an active absorber in solar cells, as it also possesses a high absorption coefficient --above 10$^{5}$ cm$^{-1}$-- for energies greater than 1.95 eV \cite{ferrer2013optical}.
In addition, It has been experimentally shown that the in-plane elasticity modulus of TiS$_3$ thin films (samples thicker than 100 nm) vary between 15 and 26 GPa \cite{liu2023unexpected}, which are comparable to the values of perovskite thin films currently used in flexible solar cells (14--35 GPa) \cite{li2023progress}.
Furthermore, Baraghani \emph{et al.} \cite{baraghani2021printed} studied the carrier transport and spectral noise density in printed electronic devices based on TiS$_3$ inks, finding evidence that these thin films can potentially be used in printed electronics. This study is particularly interesting as it points towards the roll-to-roll manufacturing of next-generation flexible solar cells, which are highly attractive due to their low cost and high-throughput mass production \cite{weerasinghe2024first}. Thus the combination of material abundancy, nontoxicity, appropriate band gap, high absorption coefficient, and felxibility render TiS$_3$ thin films a promising alternative for applications in flexible photovoltaics. In that respect, however, a deeper microscopic understanding is required. 

Herein, we use  many-body perturbation theory (MBPT) based on \emph{ab-initio} density functional calculations (DFT), which give accurate prediction of several microscopic properties including the excitonic and optical ones  \cite{onida2002electronic,bernardi2015theory,villegas2022efficient}. The results indicate that \ce{TiS3} has an electronic band gap of $\sim$1.1 eV, and possesses a high absorption coefficient, which is in good agreement with available experimental measurements \cite{suk2023polarization}. We then combine this method with the transfer matrix approach to investigate the optical and photovoltaic response of \ce{TiS3} thin films for single-junction and tandem solar cells. 
The photovoltaic potential of both devices are evaluated by studying the carrier generation rate and the idealized external quantum efficiency over a broad energy range that includes the near-infrared and visible spectra. The optimized single-junction solar cell device provides a ${\sim}$26.6$\%$ maximum PCE. In addition, our proposed optimized tandem cell composed of \ce{TiS3} and perovskite thin films delivers a maximum PCE of approximately 32.8\% for an overall active layer thickness of $\sim$1.3-$\mu$m.
\section{Methods}
\subsection{Quantum material simulation}
First-principles calculations based on density functional theory (DFT) as implemented in the \textsc{quantum} \textsc{espresso} package \cite{giannozzi2009quantum} were carried out to study the charge density of the system's ground state. The Perdew-Burke-Ernzerhof generalized-gradient approximation is employed to describe the exchange-correlation functional along with a kinetic energy cutoff of 90 Ry, and a Monkhorst-Pack \textbf{k}-point sampling of 10 $\times$ 8 $\times$ 5. The structures were fully relaxed to their equilibrium positions with residual forces smaller than 0.01 eV/\AA\ and pressures on the lattice unit cell smaller than 0.08 kbar. Van der Waals corrections within the semi-empirical dispersion scheme (PBE-D2) as proposed by Grimme were also employed \cite{grimme2006semiempirical}. The optimized lattice parameters were $a$=4.987, $b$=3.40, and $c$=8.89 $\AA$, which agree well with previous reported experimental and theoretical values \cite{furuseth10crystal,abdulsalam2016electronic}.

In order to correct the band gap, the quasiparticle energies of bulk TiS$_3$ were calculated by using the MBPT within the single-shot GW approximation (G$_0$W$_0$) as implemented in the \textsc{yambo} code \cite{yambo}. The dielectric screening was computed on a 8 $\times$ 6 $\times$ 4 \textbf{k}-grid within the Plasmon-pole approximation considering an energy cutoff of 80 Ry for the exchange potential, and 14 Ry for the screening potential $W_0$. The dynamical dielectric screening and the self-energy were computed including 800 and 240 bands, respectively. The excitonic optical spectra were computed by solving the Bethe-Salpeter equation (BSE) within the Tamm-Dancoff approximation. We took advantage of the double-grid approach \cite{kammerlander2012speeding}, for which the kernel matrix elements are first calculated on a coarse \textbf{k}-grid (the same employed for GW corrections) and then interpolated on a finer \textbf{k}-grid of 24 $\times$ 16 $\times$ 12 with six valence bands and seven conduction bands. The excitonic response is characterized by the imaginary part of the macroscopic dielectric function,
\begin{equation}
	\varepsilon_{2}(\omega) = \frac{8\pi e^2}{N_{k}\Omega}\sum_{S}
	\left|\mu_{S}\right|^{2}\delta\left(\omega-E_{S}\right),
	\label{eq1}
\end{equation}
where $\mu_{S}=\sum_{v c \mathbf{k}} A_{vc\textbf{k}}^{S}\cdot\left<v\mathbf{k}|\textbf{r}|c\mathbf{k}\right>$  are the excitonic transition dipoles. Here $A_{vc\textbf{k}}^{S}$ and $\textbf{r}$ are the excitonic wavefunction for the S-th state, and position operator, respectively. $\Omega$ represents the real space unit cell volume, and $N_k$ the
number of points in the Brillouin zone sampling. We considered light polarized along the main crystallographic directions and an artificial broadening of 40 meV to smear out the absorption curves, which has been successfully used to  
accurately reproduce the optical properties of bulk transition metal dichalcogenides \cite{villegas2024optical}. The convergence tests indicate that the set of GW-BSE parameters mentioned above allow us to accurately describe the optical properties within an error threshold of about 60 meV (see the supplemental material).
\subsection{Optical and photovoltaic device simulations}
The optical simulations, which enable us to monitor the electric field $\textbf{E}$ in each layer, are implemented by solving the Maxwell's equations within the transfer matrix method \cite{ohta1990matrix}. Thus, the optical absorptance in each layer of the device is computed by
\begin{equation}
	A(\omega)=\int n(\omega) \alpha (\omega) |\textbf{E}(\omega,z)|^{2}\,\dd z,
	\label{eq2}
\end{equation}
where $n(\omega)$ and $\alpha(\omega)=4\pi \kappa(\omega)/\lambda$ are the real part of the refractive index and absorption coefficient respectively, the latter being related with the imaginary part of the refractive index $\kappa(\omega)$. Note that the incident electric field must be connected to the solar irradiance \cite{islam2021depth,pettersson1999modeling}. 
Considering that all the absorbed light is used to excite carriers, one can obtain the carrier generation rate in the device
\begin{equation}
	\Delta G(\omega,z) = \hbar \, \phi_{\text{sun}}(\omega) \, n(\omega)\, \alpha (\omega) \,|\textbf{E}(\omega,z)|^{2} \Delta \omega.
	\label{eq3}
\end{equation}
Here $\phi_{\text{sun}}(\omega)$ is the photon flux density associated to the AM1.5G solar spectral irradiance \cite{nrelam1.5}. The photo-current density generated at a given layer is determined by
\begin{equation}
	J_{\text{sc}}=q \hbar \int A(\omega) \phi_{\text{sun}}(\omega)\, \dd\omega,
	\label{eq4}
\end{equation}
where $q$ represents the electron charge. Note that Eq. (\ref{eq4}) assumes ideal carrier collection, implying that each absorbed photon generates a single electron-hole pair, which ultimately contributes to photocurrent generation. Following the optimization recipe proposed by Jost \emph{et al.} \cite{jost2022perovskite}, the thickness of the non-active layers (HTL, ETL and back metal) are ﬁxed to their minimum experimentally achievable value for the correct working of the tandem cell. Thus, we avoid  their optimization process to save computational time.
The PCE is calculated by the well-known relation 
\begin{equation}
	\text{PCE}=\frac{\text{max}(J\cdot V)}{P_{\text{in}}}=\frac{FF\times J_{\text{sc}} \times V_\text{oc}}{P_\text{in}},
	\label{eq5}
\end{equation}
where $FF$ is the fill factor, $V_\text{oc}$, and $P_\text{in}=100$ mW/cm$^{2}$ are the open circuit voltage and incident power density from the AM1.5G solar spectrum, respectively. The current-voltage is obtained by the ideal diode relation,
\begin{equation}
	J= J_\text{sc}-J_\text{r}^\text{total}(e^{\frac{qV}{k_\text{B}T}}-1).
	\label{eq6}
\end{equation}
Here $k_\text{B}$ and $T$ represent the Boltzmann constant, and the temperature, respectively.
The open-cirtuit voltage, defined as the voltage at which Eq. (\ref{eq6}) vanishes, can be written as
\begin{equation}
	V_{oc}=\frac{k_\text{B}T}{q}ln\left[\frac{J_{sc}}{J_{r}^{total}}+1\right].
	\label{eq6.1}
\end{equation}
Following the approach proposed by Blank \emph{et al.}  \cite{blank2017selection}, $J_\text{r}^\text{total}=J_\text{r}^\text{rad}+J_\text{r}^\text{nrad}$ represents the recombination current density, which is composed of the radiative and non-radiative contribution, respectively. The radiative contribution can be expressed as
\begin{equation}
	J_\text{r}^\text{rad}= q \pi \hbar \int_{0}^{\infty} A(\omega) \phi_\text{bb}(\omega,T)\, \dd\omega,
	\label{eq7}
\end{equation}
where $\phi_\text{bb}(\omega,T)$ is the photon flux density related to the black-body spectrum,
\begin{equation}
	\phi_\text{bb}(\omega,T)=\frac{\omega^2}{2\pi^2 h c^{2}[e^{\frac{\hbar \omega}{k_\text{B}T}}-1]}.
	\label{eq8}
\end{equation}
In addition, the non-radiative current density is expressed  as
\begin{equation}
	J_\text{r}^\text{nrad}=q \int_{0}^{d} \frac{(1-Q)}{Q}R^{rad}\,\dd x 
	\label{eq9}
\end{equation}
where $Q$ is an adjustable parameter that represents the internal luminescence quantum yield. The radiative recombination rate, considering a position-independent refractive index is obtained by the van Roosbroeck-Shockley equation \cite{van1954photon},
\begin{equation}
	R^\text{rad}=4 \pi \hbar \int_{0}^{\infty} n^{2}(\omega) \alpha(\omega) \phi_\text{bb}(\omega)\,\dd\omega.
	\label{eq10}
\end{equation}
Finally, the monolithic multi-junction solar cell is modeled by considering the series connection of the two active subcells. This design constrains the photo-current density passing through the device to the minimum value between the highest current density achieved by any of the cells, \emph{i.e}, $J=\text{min}[J_\text{bottom}(V_\text{max}^\text{bottom}), J_\text{top}(V_\text{max}^\text{top})]$, while the voltage across the tandem cell is obtained by adding up the maximum voltage achieved in each sub cell,  $V=V_\text{max}^\text{bottom}+V_\text{max}^\text{top}$. Hence, the power conversion efficiency for the tandem device is given by \cite{alharbi2015theoretical},
\begin{equation}
	\text{PCE}^\text{T}=\frac{J \cdot \sum_{i} V_{i}}{P_\text{in}}.
\end{equation}
In all simulations the series resistance is neglected.
\begin{figure*}[t!]
	\centering
	\includegraphics[width=1.0\linewidth]{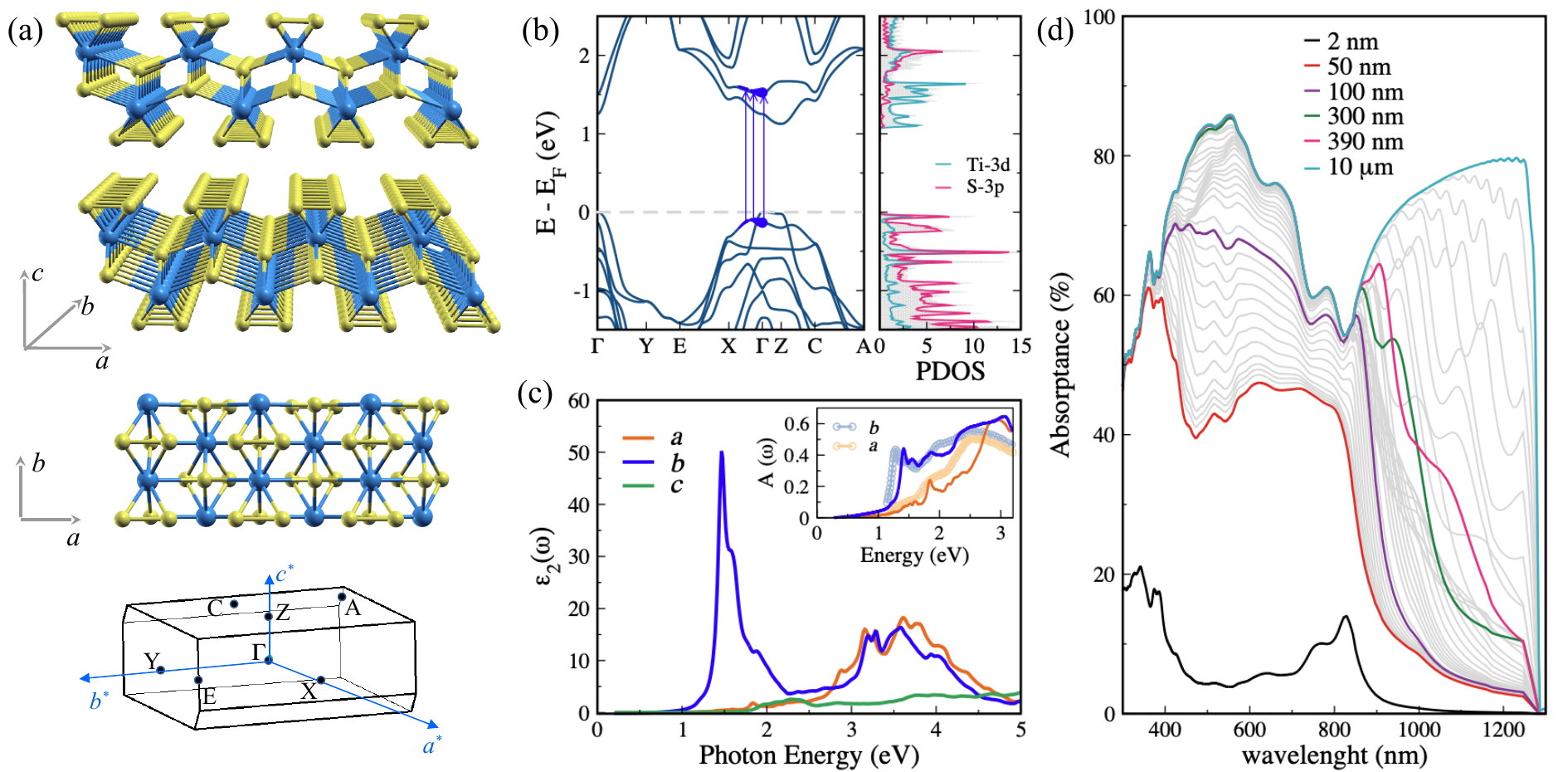}
	\caption{(a) Crystal structure of bulk TiS$_3$ and its corresponding Brillouin zone showing the high-symmetry points. (b) Quasiparticle electronic structure along the main symmetry points and the corresponding total and projected density of states. The vertical arrows indicate the crystallographic position of the optical transitions that originate the most prominent resonance in the optical response. (c) Imaginary part of the dielectric function for different polarizations of light. The inset shows a comparison of the calculated optical absorptance  with experimental measurements \cite{suk2023polarization}. The theoretical absorptance is obtained using Fresnel's equations, assuming a 60 nm thick TiS$_3$ layer on a SiO$_2$ substrate. (d) Absorptance of TiS$_3$ for thicknesses ranging from 2 nm to 10 $\mu$m. Here, we assume that the active layer is deposited on a semi-infinite SiO$_2$ substrate following the analytical model proposed in a previous work \cite{villegas2024elucidating}.}
	\label{fig1}
\end{figure*}
\section{Results and discussion}
\subsection{Electronic and Optical properties of  bulk TiS$_3$}
Bulk TiS$_3$ is a monoclinic crystal with space group P2$_{1}$/\emph{m}. It is composed of non-planar layers stacked along the $c$-direction and kept together by van der Waals interactions, which according to our simulations possesses a characteristic interlayer distance of $\sim$3.15 $\AA$. Each layer consists of atomic chains with strong covalent bonds extending along the $b$-direction, in which titanium and sulfur atoms form trigonal prisms (see Fig.~\ref{fig1}{\color{blue}{a}}).

The quasiparticle electronic structure obtained within the $G_0W_0$ approach is shown in Fig.~\ref{fig1}{\color{blue}{b}}.
The system presents a fundamental electronic band gap of $\sim$1.12 eV with dipole-forbidden transitions at the $Z$-point due to inversion symmetry of the wavefunction at the band edges, similar to what occurs for the single layer case \cite{suk2023polarization,wang2021high}. Our calculated electronic band gap is consistent with previous scanning tunneling spectroscopy measurements that report an electronic band gap of $\sim$1.2 eV \cite{molina2015electronic}. The projected density of states reveals that the band edge states are composed mainly of hybridized Ti-3$p$ and S-3$p$ states. 

The excitonic optical spectrum, obtained by solving the Bethe-Salpeter equation (BSE), presents a highly anisotropic photoresponse between 1.1 and 2.5 eV. In Fig.~\ref{fig1}{\color{blue}{c}} we present the imaginary part of the dynamical dielectric function spanning the infrared to near-ultraviolet spectrum for different crystallographic directions (see the supplemental material for the real part of the dielectric function). In particular, for light polarized along the $b$-direction a very intense absorption onset at 1.49 eV is observed. This peak results from dipole-allowed transitions occurring along the $\Gamma$-$X$ path, as depicted by arrows in Fig.~\ref{fig1}{\color{blue}{b}}. In contrast, along the $a$ and $c$ directions, the material is almost transparent in the near-infrared range. Moreover, between 2.5--5 eV, only the in-plane polarized responses exhibit pronounced peaks.
\begin{figure*}[ht!]
	\centering
	\includegraphics[width=1.0\linewidth]{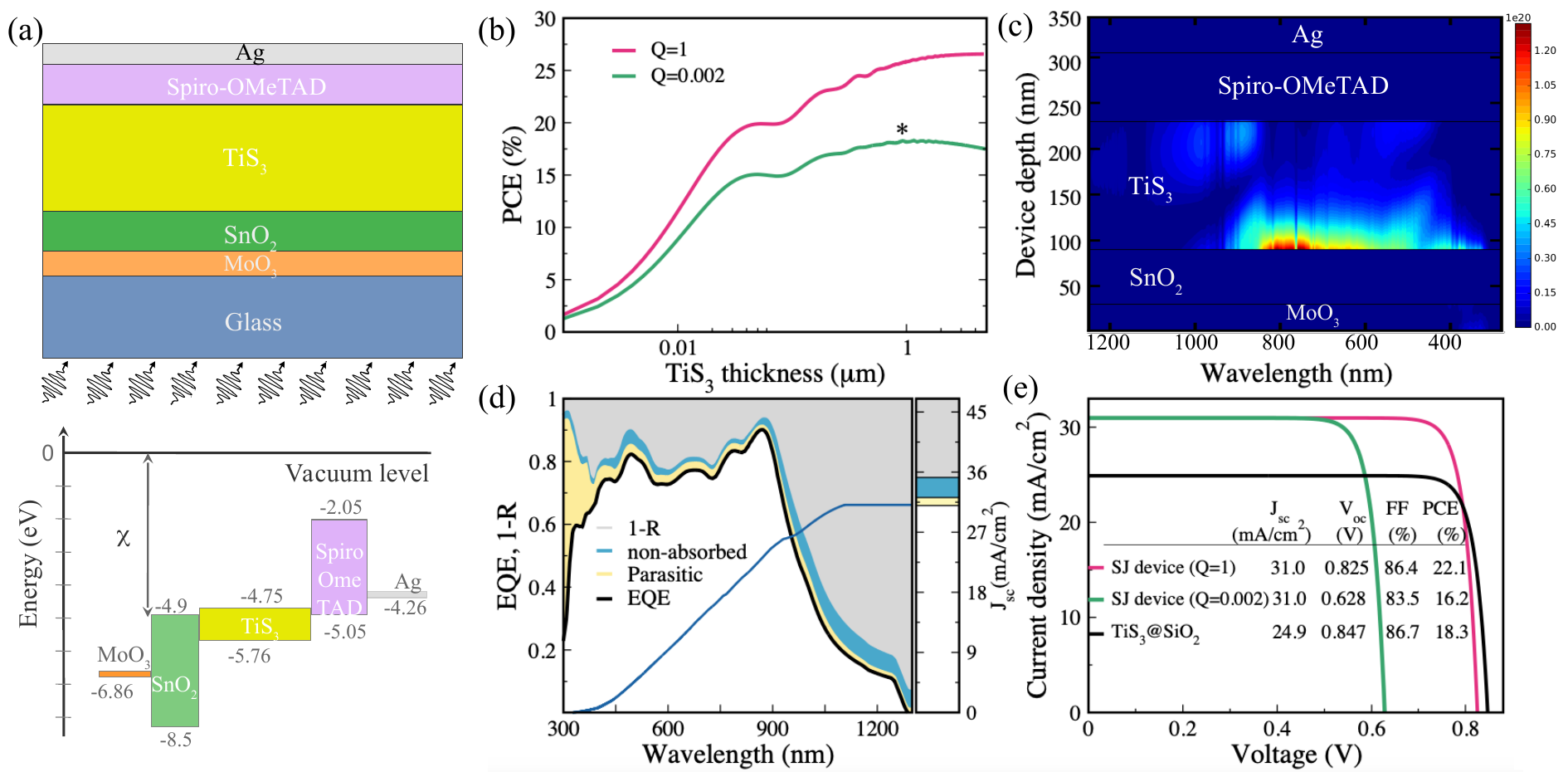}
	\caption{(a) Top panel: architecture of the proposed single-junction solar cell. Bottom panel: band alignment of each layer constituting the device. (b) Power conversion efficiency as a function of thickness in the radiative  (pink line) and non-radiative (green line) limit. The 
		asterisk indicates the highest value for the PCE. (c) Carrier generation rate versus the device depth and light wavelength. (d) Optical losses as a function of wavelength. The white area represents the absorptance relative to the active layer while the grey, yellow and cyan regions correspond to the reflection, parasitic and absorption losses. (e) Current density-voltage relative to the device without (pink curve) and with (green curve) non-radiative recombination effects. For completeness, the current-voltage curve for TiS$_3$ on SiO$_2$ is shown (black curve). The values for the band alignments values for Ag, MoO$_3$, Spiro-OMeTAD and SnO$_2$ were taken from existing literature \cite{michaelson1977work,kroger2009role,leijtens2012hole,chizhov2016visible}}
	\label{fig2}
\end{figure*}

To gain insight into the relevance of excitonic effects, we compare the optical spectra using the BSE with the one obtained at the independent quasiparticle level (see supplemental material). There, we note significant differences in the intensity over the  whole spectrum, which highlights the importance of excitonic effects. In particular, we predict the presence of low-energy excitons with binding energies of  $\sim$110 meV, which is in excellent agreement with previous theoretical and experimental studies \cite{molina2015electronic}.  
This value is larger than those of Si and CdTe, whose exciton binding energies are approximately 15 meV \cite{green2015,strzkoski2013}. Since these values are lower than the thermal energy at room temperature, excitons in these materials are expected to readily dissociate into free electrons and holes, which can act as free charge carriers. Despite this, excitonic effects in these low-exciton-binding-energy materials can still have a slight impact on photovoltaic performance \cite{zhang1998effects}.
On the other hand, excitonic effects in \ce{Cu2O} and \ce{CsPbCl3}, which have exciton binding energies of $\sim$100 meV \cite{steinhauer2020} and 75 meV \cite{protesescu2015}, respectively (comparable to that of \ce{TiS3}), are expected to play a key role in the performance of photovoltaic devices. For instance, it has been demonstrated that, due to the large exciton binding energy in \ce{Cu2O}, approximately one-third of photogenerated carriers originate from excitons. These high exciton densities also contribute to enhanced current collection at energies below the electronic band gap, leading to an efficiency increase of up to 2\% \cite{omelchenko2017excitonic}.

To validate the overall accuracy of our approach, the inset of Fig.~\ref{fig1}{\color{blue}{c}} compares a recent experimental measurement of absorptance  (light dotted lines) with our theoretical calculations for a 60 nm thin-film. We can clearly see that our results are in good agreement with the experimental measurements and capture the main features and profile of the optical response for both light polarizations. We also calculate the absorption coefficient of  TiS$_3$, which presents values higher than 10$^{5}$ cm$^{-1}$ below 900 nm. This  behavior persist throughout the entire visible range being, in some cases, considerably larger than the absorption coefficient of Si and MAPbI$_3$ (see supplementary material). 

In Fig. \ref{fig1}{\color{blue}{d}}, we assess the performance of \ce{TiS3} as an active absorber layer by varying its thickness from 2 nm (corresponding to roughly a three-layer system) to 10 $\mu$m. 
The thickness-dependence of the absorptance shows a maximum of $\sim$85$\%$ at $\lambda$ $\sim$560 nm that can be obtained with a sample of only 300-nm-thick. This value is enough to saturate the absorptance between 300 and 854 nm. The enhancement of absorptance beyond 854 nm requires samples thicker than 1$\mu$m. 
\subsection{Single-junction solar cell}
Combining the information about the isolated bulk material with other ingredients, we can propose a realistic device.
The lateral view of a prototypical \ce{TiS3} single junction solar cell is schematically represented in Fig.~\ref{fig2}{\color{blue}{a}}. The architecture is formed by five layers covered by a glass, where the active layer is sandwiched between a hole transport layer (HTL) and an electron transport layer (ETL) made of 75-nm-thick Spiro-OmeTAD, and 60-nm-thick \ce{SnO2}, respectively. These layers aid in the collection of excited carriers formed in \ce{TiS3}. The design also includes a transparent conductive layer (TCL) made of \ce{MoO3} with thickness of 30 nm  and a 40-nm-thick silver layer that enhances the absorption in the active layer as it reflects back the light. The TCL, ETL, HTL, and back reflector materials were carefully chosen to facilitate the proper band alignment with \ce{TiS3}. Here we employ an experimentally estimated electron affinity ($\chi$) for \ce{TiS3} of $\sim$4.75 eV \cite{agarwal2018anomalous}, while the electron affinities and work function of other layers were taken from previous works \cite{michaelson1977work,kroger2009role,leijtens2012hole,chizhov2016visible}. 
The band alignments, depicted at the bottom of Fig.~\ref{fig2}{\color{blue}{a}}, show that the conduction band of \ce{TiS3} is 0.15 eV higher than that of the \ce{SnO2} layer. This alignment may facilitate the transport of photo-excited electrons towards the \ce{MoO3} layer. In addition, the valence band offset of about 0.7 eV between \ce{TiS3} and Spiro-OmeTAD enables hole transport to the metal electrode. On the other hand, the high band offset at the interfaces of \ce{TiS3}/\ce{SnO2} in the valence band, and \ce{TiS3}/Spiro-OmeTAD in the conduction band block the hole and electron transport towards the transparent and metallic electrode, respectively. 

An optimized thin-film device should be designed encountering a balance between thickness and efficiency, given that, in practice, thicker films possess a larger amount of defects, which introduce non-radiative recombination centers that reduce the overall device PCE \cite{luo2020minimizing,yamaguchi2023overview}. Hence, following the model proposed by Blank \emph{et al.} \cite{blank2017selection} (see methods section), we estimate the PCE including non-radiative recombination losses through a tunable factor $Q$, which is closely related to the internal luminescence yield. Fig.~\ref{fig2}{\color{blue}{b}} shows the dependence of PCE on the active layer thickness. In the radiative limit there is no recombination and $Q=$1. In this case, we predict that active layers exceeding 140 nm could yield PCEs higher than 22$\%$, besides, PCE saturates to 26.6$\%$ for thickness larger than $\sim$3$\mu$m. 

We then consider a more realistic scenario that includes non-radiative losses and estimate a lower limit for PCEs by choosing $Q=$0.002 as a representative value. This value is one order of magnitude smaller than the losses typically found in Si (0.017) \cite{yoshikawa2017silicon} and Perovskites (0.07) \cite{jiang2019surface}. In this regime, our calculations show that a 140-nm-thick active layer delivers a PCEs of 16.2$\%$. By increasing the thickness to 980 nm, the efficiency is slightly enhanced by 2$\%$ and then decreases due to non-radiative scattering events. In order to simulate a practical single-junction solar cell device, based on these results, we adopt a thickness of 140 nm for the active layer. This value resembles the estimated experimental value for the exciton diffusion length in \ce{TiS3} thin flims, which has been reported to be $\sim$130 nm \cite{cui2016time}.

Fig.~\ref{fig2}{\color{blue}{c}} presents the carrier generation rate as a function of the  light wavelength (horizontal-axis) and device depth (vertical-axis), calculated using Eq. (\ref{eq3}).  Electron-hole pairs are efficiently generated at the SnO$_2$/TiS$_3$ interface for photon wavelengths between 700--830 nm. Interestingly, a non-vanishing generation rate between 900--1120 nm is also observed over the whole active layer. This is due to the low absorption coefficient of \ce{TiS3} in this range, which ultimately allows photons to travel deeper into the active layer before being absorbed.

The optical losses and  ideal external quantum efficiency (EQE) are shown in Fig.~\ref{fig2}{\color{blue}{d}}. Here, the EQE accounts for real absorptivity and assumes ideal collection of photogenerated carriers. The reflected photons at the glass-air interface, the photons absorbed by the TCL, HTL, and ETL layers, and the non-absorbed photons due to the limited thickness of the device are categorized as reflection, parasitic, and absorption losses, respectively. 
The results indicate that an active layer of 140-nm-thick absorbs nearly 66\% of incident photons to photo-generated carriers. Moreover, 25$\%$ of incoming photons are lost due to reflection, while parasitic losses represent 2.5$\%$ and dominate the short-wavelength region of the spectrum. The remaining 6.5$\%$ corresponds to absorption losses. We note that increasing the thickness to 390 nm reduces reflection losses to 18$\%$, yielding about 78$\%$ EQE (see supplemental material).

In order to understand the effect of the back reflector, in Fig.~\ref{fig2}{\color{blue}{e}}, we compare the current density-voltage curve results for the single junction (SJ) with that of TiS$_3$ deposited only on SiO$_2$ (TiS$_3$@SiO$_2$), that consists simply of three layers: air, TiS$_3$, and SiO$_2$. The main difference lies in the current density enhancement for the SJ device, reaching $\sim$31 mA/cm$^{2}$ compared to 24.9 mA/cm$^{2}$ for TiS$_3$@SiO$_2$. This increase is primarily ascribed to the presence of the metal back reflector that allows the active layer to reabsorb additional photons and the interference processes occurring at the interfaces of the device. Since, the fill factors and open-circuit voltages are only slightly modified, we argue that the change in J$_{sc}$ is responsible for the PCE increase from 18.3$\%$ in TiS$_3$@SiO$_2$ to 22.1 $\%$ in the SJ device. It should be noted, as seen from eq. (\ref{eq6.1}), that the enhancement or decrease of V$_{oc}$ ultimately depends on the ratio $J_{sc}/J_{r}^{tot}$. Hence, the enhancement of $J_{sc}$ does not necessarily imply an increase in V$_{oc}$, as occurs in the SJ-device (Q = 1) case. This can be attributed to the fact that the rate of increase for $J_{sc}$ differs from that of $J_{r}^{tot}$, since the latter is typically thousands of times smaller than J$_{sc}$ \cite{bercx2016first}. This explains the superior value of V$_{oc}$ for the device based on TiS$_3$/SiO$_2$.
\begin{figure*}[ht!]
	\centering
	\includegraphics[width=1.0\linewidth]{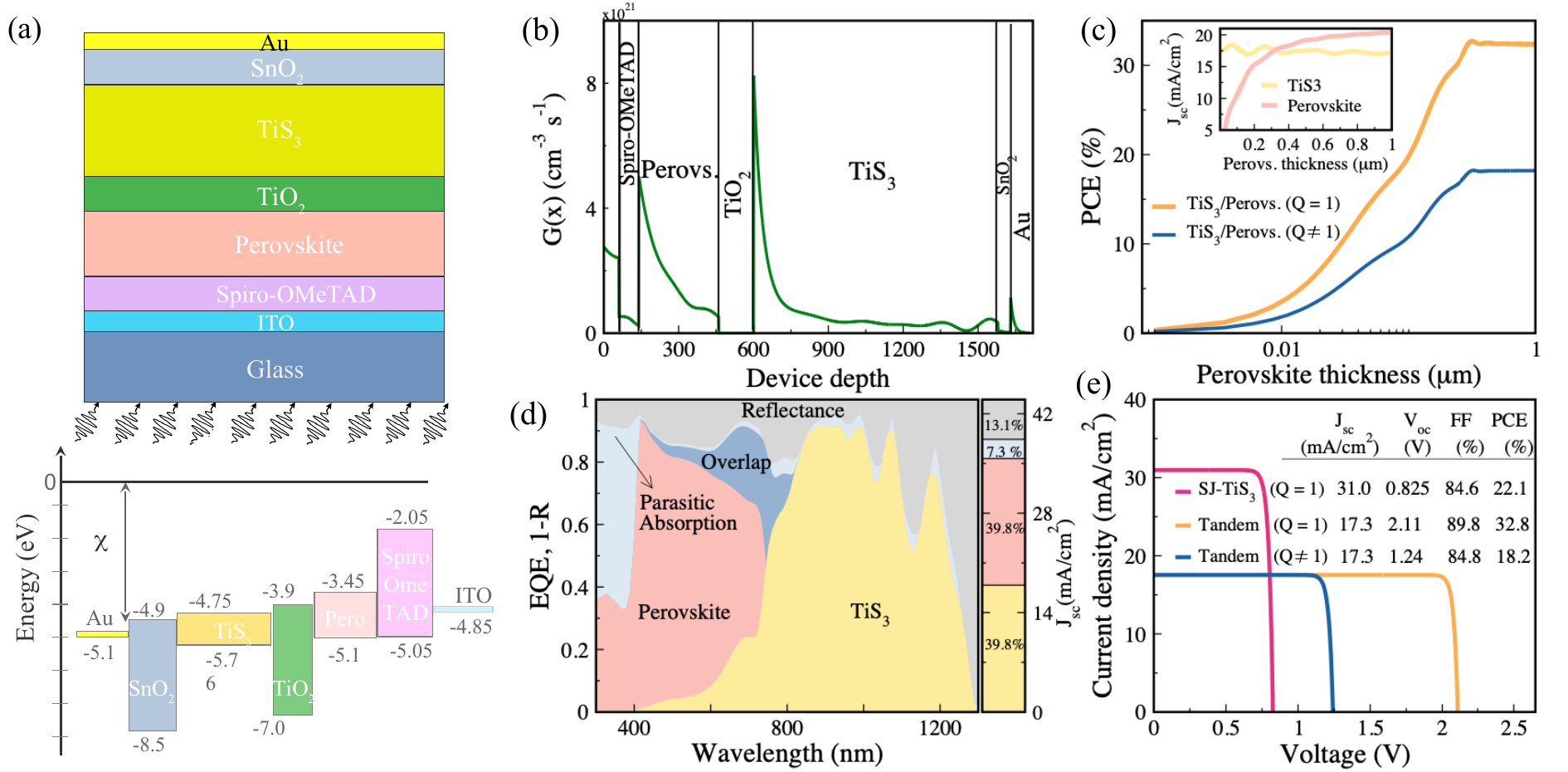}
	\caption{(a) Proposed tandem solar cell device and band alignments constructed from the electron affinity of each layer. (b) Integrated carrier generation rate versus the device depth. (c) Power conversion efficiency as a function of perovskite thickness with (blue curve) and without (orange curve) non-radiative recombination losses. The inset shows $J_{sc}$ in each active layer when the perovskite thickness is varied. (d) Optical gain and losses as a function of wavelength. The yellow and pastel pink areas represent the region of the active layers. The reflectance and parasitic losses are indicated by gray and sky-blue. (e) Current density versus voltage of the tandem device with (blue curve) and without (orange curve) recombination losses. For comparison purposes, the curve for the SJ device (pink curve) is also shown. The band alignment for \ce{MAPbI3}, and ITO are taken from previous works \cite{caputo2019electronic,cai2020ultraviolet}.}
	\label{fig3}
\end{figure*}

A comparison of the SJ device in the radiative limit with the one that takes into account non-radiative recombination processes (Q=0.002) highlights the reduction of the open-circuit voltage by 0.2 V. This leads to a decrease in PCE to 16.2$\%$. This result suggests that with only 140-nm-thick TiS$_3$, achievable efficiencies are close to those of well established thin-film devices based on CIGS, which typically employ active layers of few micrometers \cite{green2023solar} and would have large problems involving surface reconstruction for thinner samples.
\subsection{Tandem solar cell}
To bring PCEs to the next level, we investigate the feasibility of using TiS$_3$ thin-films as the bottom cell in tandem architectures. 
The proposed architecture for the monolithic tandem solar cell and the relevant band alignments are depicted in Fig.~\ref{fig3}{\color{blue}{a}}. The device is constituted by layers of ITO, Spiro-OMeTAD, ­Perovskite, TiO$_2$, TiS$_3$, SnO$_2$ and Au with the thicknesses of 60, 80,
320, 140, 980, and 80 nm, respectively.  We consider the perovskite layer (\ce{MAPbI3}) as the top layer due to its key role in current highly efficient thin-film solar cell devices.

The integrated carrier generation profile is presented in the inset of Fig.~\ref{fig3}{\color{blue}{b}}. There it can be seen that carriers are efficiently generated at the Spiro-OMeTAD/Perovskite and TiO$_2$/TiS$_3$ interfaces. In both active layers, the carriers diffuse through the entire length of both layers with non-vanishing values. Moreover, we note a slightly higher intensity in the perovskite layer as photons travel deeper into the device. It is also important to mention that the TiO$_2$ and SnO$_2$ layers do not generate carriers, while the ITO layer generates a significant amount of them. 

In order to maximize the power conversion efficiency of the device, we initially assess the current matching in the top and bottom active layers. In the inset of Fig.~\ref{fig3}{\color{blue}{c}} we present the current density of the active layers (with a fixed thickness of 980 nm for \ce{TiS3}) as a function of perovskite layer thicknesses. The results show that a 320-nm-thick perovskite layer is enough to achieve current matching. Moreover, the PCE of the tandem device as a function of perovskite thickness (see Fig.~\ref{fig3}{\color{blue}{c}}) also indicates, as expected, that $\sim$320 nm thick perovskite layer provides the maximum achievable PCE of about 32.8$\%$ for the radiative case ($Q$=1) and $\sim$18.2$\%$ for the more realistic case where non-radiative recombination is considered ($Q_{\text{\ce{TiS3}}}$=0.002 and $Q_{\text{Perovs}}$=0.08). 

Examining the ideal external quantum efficiency of the tandem device, presented in Fig.~\ref{fig3}{\color{blue}{d}} we note that the active layers absorb nearly $\sim$80$\%$ of incoming photons, while $\sim$7.3$\%$ and 13.1\% are lost to parasitic and reflectance losses. 
The current density versus voltage curves for the tandem device in the radiative ($Q$=1) and non-radiative limits ($Q\neq$1) are shown in  Fig.~\ref{fig3}{\color{blue}{e}}. There, it is clear that the open-circuit voltage V$_{oc}$ and the fill factor (FF) decrease by  0.87 V and  5$\%$, respectively when we include non-radiative recombination losses. 
It is worth mentioning that if the recombination losses are ignored, the tandem device outperforms the studied SJ device by almost 33$\%$. 

\subsection{Future directions and limitations}
Our prototype device with wafer thickness smaller than 1$\mu$m makes the design of flexible and wearable devices possible. This is particularly relevant in light of a very recent study that reported the design of truly flexible silicon solar cells with certified efficiencies of 26.06 $\%$ and a wafer thickness of 57 $\mu$m \cite{li2024flexible}. This demonstrates that wafer samples thinner than 100 $\mu$m  can be effectively bent with curvature radius smaller than 38 mm, allowing for a high degree of flexibility. 

We also mention that the PCE of these prototypical devices can be further enhanced by designing resonant structures that combine supercell gratings with waveguide slabs, as demonstrated in perovskite solar cells \cite{feng2023resonant}. Due to their transparency and unique propagation properties \cite{rickhaus2013ballistic,zhang2009guided,villegas2010comment,villegas2012}, we argue that graphene-based waveguides can be appealing candidates for integration into TiS$_3$ resonant solar cells. Regarding the tandem devices, future research efforts could explore new scenarios in which TiS$_3$ is combined with other materials that have a high absorption coefficient and suitable band gaps in the range of 1.6 to 1.7 eV.

Since, in real devices, the reduction of V$_{oc}$—which prevents the efficient extraction of photo-excited carriers \cite{luo2020minimizing}—is related to electronic losses due to non-radiative recombination processes such as electron-electron, electron-phonon, or carrier-impurity scattering pathways, we argue that performing simulations incorporating the phenomenological parameter $Q$ is crucial for estimating a lower limit for the PCE based on reasonable assumptions.

Although the results obtained for the prototypical TiS$_3$ photovoltaic devices rely on 1D modeling, which assumes perfectly planar layers and spatially uniform voltages and current densities for each layer in the cell, we argue that these assumptions are reasonable whenever lateral current transport within the cell layers is negligible. However, if the experimental realization of the device reveals the significance of lateral currents, a three-dimensional model should be used to account for non-uniform conditions. 
Finally, despite the existence of such currents, several experimental challenges must still be addressed before TiS$_3$ solar cells become a reality, including increasing the exciton diffusion length, detailed characterization of bending radius, and compatibility with printed electronics. Nevertheless, we believe that our findings can pave the way towards the experimental realization of next-generation flexible solar cells based on TiS$_{3}$.

\section{Conclusions}
In summary, we employed excited-state density functional theory calculations along with the transfer matrix approach to investigate the optical and photovoltaic response of TiS$_3$ thin-film single-junction and tandem solar cells. Overall, our results highlight important features that make TiS$_3$ thin films a promising semiconductor for designing flexible solar cells.
First, our first-principles simulations suggest that TiS$_{3}$ possesses a high absorption coefficient in the visible range and excitons with binding energies of approximately 110 meV, making them robust against thermal fluctuations and potentially contributing to photocurrent generation. Contrary to previously available measurements of optical constants, our simulations enable us to map out the infrared to ultraviolet-visible range of the electromagnetic spectrum, where experimental measurements either have low resolution or are unavailable.
Additionally, TiS$_{3}$ exhibits a low absorption coefficient only in the near-infrared range, allowing photons to penetrate deeply into the material. This characteristic potentially facilitates carrier separation at both the electron and hole transport layers. While the exciton diffusion length for a 3-$\mu$m-thick TiS${3}$ layer has been reported to be around 130 nm, our predicted $\sim$16.2$\%$ PCE, for a device of similar thickness and accounting for non-radiative recombination losses, can be considered a reasonable lower limit for single-junction device efficiency. Moreover, our calculations indicate that an optimized single-junction solar cell can achieve a maximum PCE of ${\sim}$22.1$\%$ with only a 140-nm-thick active layer, which can be further enhanced to 25.8$\%$ for a 1-$\mu$m-thick active layer.
Finally, the tandem cell design, which combines \ce{TiS3} thin films with a perovskite layer, achieves a maximum PCE of 32.8$\%$ when considering a total active layer thickness of approximately 1.3-$\mu$m. Therefore our results suggest the potential of \ce{TiS3} thin films for the design of highly efficient, flexible photovoltaic devices. At the same time, the protocol presented here can be applied to different materials with minimal parameter input while providing a microscopic description that accounts for excitonic effects.


\section{Acknowledgements}
E.M.Jr acknowledges the financial support from the Brazilian agency FAPESP, Grant No. 20/13172-8. ARR acknowledges support from FAPESP (Grant No. 2017/02317-2 and 2021/14335-0). PV thanks FAPESP
for a visiting professor fellowship (FAPESP Grant No. 2021/13720-8). This research was supported by resources supplied by CENAPAD-SP, LNCC-Santos Dumont, and the Center for Scientific Computing (NCC/GridUNESP) of the UNESP.
 

\end{document}